\documentstyle[seceq,epsf,preprint,mbf,wrapft]{ptptex}

\pubinfo{}  
\preprintnumber[\textwidth]{\begin{flushright}
ANL-HEP-CP-00-001 \\
DESY 00-007\end{flushright}}

\title{
Longitudinal Spin Dependence of Massive Lepton Pair Production
}

\author{
Edmond L. {\sc Berger} $^{a)}$ \footnote{E-mail address: berger@anl.gov}, 
Lionel E. {\sc Gordon} $^{b,c)}$, and {Michael {\sc Klasen} $^{d)}$}

\inst{
$^{a)}$High Energy Physics Division, Argonne National Laboratory, Argonne, IL 60439 \\
$^{b)}$Jefferson Laboratory, Newport News, VA 23606 \\
$^{c)}$Hampton University, Hampton, VA 23668 \\
$^{d)}$II.\ Inst.f.Theor.Physik, Univ.Hamburg, Luruper Chaussee 149, D-22761 Hamburg
}

}
\abst{
We show that massive lepton-pair production, the Drell-Yan process,   
should be a good source of independent constraints on the polarized gluon 
density, free from the experimental and theoretical complications of photon 
isolation that beset studies of prompt photon production.  We provide 
predictions for the spin-averaged and spin-dependent differential cross 
sections as a function of transverse momentum $Q_T$ at energies relevant for 
the Relativistic Heavy Ion Collider (RHIC) at Brookhaven.
}

\begin{document}

\maketitle


\makeatletter
\if 0\@prtstyle
\def\asp{.3em} \def\bsp{.26em}
\else
\def\asp{.3em} \def\bsp{.3em}
\fi \makeatother

\section{Introduction}

Massive lepton-pair production, $h_1 + h_2 \rightarrow \gamma^* + X;
\gamma^* \rightarrow l \bar{l}$, known as the Drell-Yan process~\cite{ref:DY},
and prompt real photon production, $h_1 + h_2 \rightarrow \gamma + X$, are two 
of the most valuable probes of short-distance behavior in hadron reactions.  
They supply critical information on parton momentum densities and 
opportunities for tests of perturbative quantum chromodynamics (QCD).
Spin-averaged parton momentum densities may be extracted from spin-averaged 
nucleon-nucleon reactions, and spin-dependent parton momentum densities from 
spin-dependent nucleon-nucleon reactions.  An ambitious experimental program 
of measurements of spin-dependence in polarized proton-proton reactions will 
begin soon at Brookhaven's Relativistic Heavy Ion Collider (RHIC) with 
kinematic coverage extending well into the regions of phase space in which 
perturbative quantum chromodynamics should yield reliable predictions.

The Drell-Yan process has tended to be thought of primarily as a source of 
information on quark densities.  Indeed, the mass and longitudinal momentum 
(or rapidity) dependences of the cross section (integrated over the transverse
momentum $Q_T$ of the pair) provide essential constraints on the 
{\it antiquark} momentum density, complementary to deep-inelastic lepton 
scattering from which one gains information of the sum of the quark and 
antiquark densities.  Prompt real photon production, on the other hand, is a 
source of essential information on the {\it gluon} momentum density.  At 
lowest order in perturbation theory, the reaction is dominated at large values 
of the transverse momentum $p_T$ of the produced photon by the ``Compton" 
subprocess, $q + g \rightarrow \gamma + q$.  This dominance is preserved at 
higher orders, indicating that the experimental inclusive cross section 
differential in $p_T$ may be used to determine the density of gluons in the 
initial hadrons.~\cite{ref:BQX}

In this paper, we summarize recent work~\cite{ref:BGKDY,ref:BGKDYX}, in which 
we demonstrate that the Compton subprocess, $q + g \rightarrow \gamma^* + q$
also dominates the Drell-Yan cross section in polarized and unpolarized 
proton-proton reactions for values of the transverse 
momentum $Q_T$ of the pair that are larger than roughly half of the pair 
mass $Q$, $Q_T > Q/2$.  The Drell-Yan process is therefore a valuable, 
heretofore overlooked, independent source of constraints on the spin-averaged 
and spin-dependent {\it gluon densities}.  
Although the Drell-Yan cross section is smaller than the prompt photon cross 
section, massive lepton pair production is cleaner theoretically since 
long-range fragmentation contributions are absent as are the experimental and
theoretical complications associated with isolation of the real 
photon.~\cite{ref:BGQ} \ Moreover, the dynamics of spin-dependence in 
hard-scattering processes is a sufficiently complex topic, and its understanding 
at an early stage in its development, that several defensible approaches for 
extracting polarized parton densities deserve to be pursued with the expectation 
that consistent results must emerge.  

\section{Unpolarized Cross Sections}
\begin{wrapfigure}{r}{6.6cm}
        \epsfxsize = 7 cm
        \centerline{\epsfbox{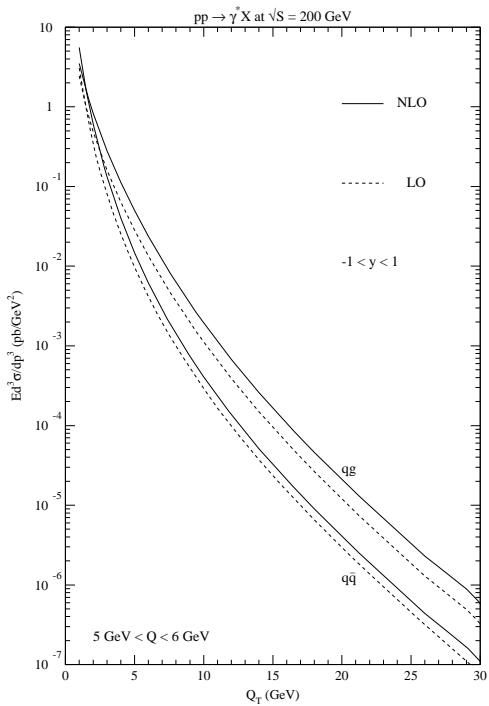}}
\caption{Lowest order and next-to-leading order perturbative 
calculations of $Ed^3\sigma/dp^3$ as a function of $Q_T$ for 
$p p \rightarrow \gamma^* X$ at $\sqrt S =$  200 GeV.  Contributions 
from the $qg$ and $q \bar{q}$ channels are shown separately.  The 
results are averaged over the rapidity 
interval - 1.0 $< y <$ 1.0 and over the interval 5.0 $<Q<$ 6.0 GeV.}
\label{fig:1}
\end{wrapfigure} 
In this Section differential cross sections for massive lepton-pair production 
are presented as functions of $Q_T$ at collider energies.  We work in the 
$\overline{\rm MS}$ renormalization scheme and set the 
renormalization and factorization scales equal.  We employ the MRST98-1 set 
of spin-averaged parton densities~\cite{ref:MRST} 
and a two-loop expression for the strong coupling strength $\alpha_s(\mu)$, 
with five flavors and appropriate threshold behavior at $\mu = m_b$.  We use 
the value $\Lambda^{(4)} = 300$ MeV of the MRST98-1 set.  The strong coupling 
strength $\alpha_s$ is evaluated at a hard scale $\mu = \sqrt{Q^2+Q_T^2}$.  
We offer results for three values of RHIC center-of-mass energy, 
$\sqrt S =$ 50, 200, and 500 GeV.     

For $\sqrt S =$ 200 GeV, we present the invariant inclusive cross section 
$Ed^3\sigma/d p^3$ as a function of $Q_T$ in Fig.~1.  Shown 
are the $q {\bar q}$ and $q g$ perturbative contributions 
to the cross section at leading order and at next-to-leading order.  For
$Q_T<$ 1.5 GeV, the $q {\bar q}$ contribution exceeds that of $q g$ channel. 
However, for values of $Q_T >$ 1.5 GeV, the $q g$ contribution becomes 
increasingly important.  The $q g$ contribution accounts for about 80 \% of 
the rate once $Q_T \simeq Q$.  Subprocesses other than those initiated by the 
$q {\bar q}$ and $q g$ initial channels contribute negligibly.

In Ref.~\citen{ref:BGKDYX}, results are shown also for a larger value of 
Q: 11 $<Q<$ 12 GeV, and for the interval 2.0 $<Q<$ 3.0 GeV.  The fractions of 
the cross sections attributable to $qg$ initiated subprocesses 
again increase with $Q_T$, growing to 80 \% for $Q_T \simeq Q$.  
In region 2.0 $<Q<$ 3.0 GeV, one would doubt the reliability of leading-twist 
perturbative descriptions of the cross section $d\sigma/dQ$, {\it integrated} 
over all $Q_T$.  However, for values of $Q_T$ that are large enough, 
$\alpha_s(Q_T)$ is small and a perturbative 
description of the $Q_T$ dependence of $d^2\sigma/dQdQ_T$ ought to be 
justified.  

\begin{wrapfigure}{r}{6.6cm}
        \epsfxsize = 7 cm
        \centerline{\epsfbox{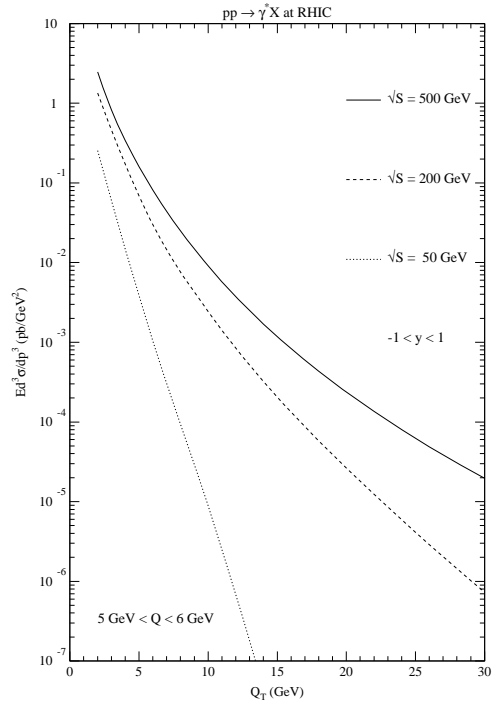}}
\caption{Differential cross sections $Ed^3\sigma/dp^3$ as a function of $Q_T$ 
for for $p p \rightarrow \gamma^* +X$ at $\sqrt S =$  
50, 200, and 500 GeV, averaged over the rapidity interval -1.0 $< y <$ 1.0 
and the mass interval 5.0 $<Q<$ 6.0 GeV.}  
\label{fig:2}
\end{wrapfigure} 
In Fig.~2, we provide next-to-leading order predictions of the differential 
cross section as a function of $Q_T$ for three values of the center-of-mass 
energy.  In order to ascertain the range of 
values of $Q_T$ that can be explored, we take  
$Ed^3\sigma/dp^3 = 10^{-3} \rm{pb/GeV}^2$ as the minimum accessible cross 
section.  This level is based on assumed luminosities~\cite{ref:bunce} of 
$8 \times 10^{31} {\rm cm}^{-2} {\rm sec}^{-1}$ at $\sqrt S =$ 200 GeV, and 
$2 \times 10^{32} {\rm cm}^{-2} {\rm sec}^{-1}$ at $\sqrt S =$ 500 GeV, along 
with runs of 10 weeks per year, equivalent to integrated luminosities of 
320 ${\rm pb}^{-1} {\rm year}^{-1}$ and 800 ${\rm pb}^{-1} {\rm year}^{-1}$.  
Better performance than anticipated of the accelerator and extended running 
time will increase the reach.  The luminosity scales roughly with $\sqrt S$
because the focusing power can be increased as the energy is reduced.  
This capability saturates near 70 GeV/c per beam, and below about 
$\sqrt S =$ 140 GeV the luminosity drops roughly as $S$.  

Adopting the nominal value \\
$Ed^3\sigma/dp^3 = 10^{-3} \rm{pb/GeV}^2$, we use 
the curves in Fig.~2 to establish that the massive 
lepton-pair cross section may be measured to $Q_T =$ 7.5, 14, and 18.5 GeV at 
$\sqrt S =$ 50, 200, and 500 GeV, respectively, when 2 $< Q <$ 3 GeV, and 
to $Q_T =$ 6, 11.5, and 15 GeV when 5 $< Q <$ 6 GeV.  In terms of reach 
in the fractional momentum $x_{gluon}$ carried by the gluon, these values of 
$Q_T$ may be converted to $x_{gluon} \simeq x_T = 2 Q_T/\sqrt S =$ 0.3, 
0.14, and 0.075 at $\sqrt S =$ 50, 200, and 500 GeV when 2 $< Q <$ 3 GeV, and 
to $x_{gluon} \simeq$ 0.24, 0.115, and 0.06  when 5 $< Q <$ 6 GeV.  On the face 
of it, the smallest value of $\sqrt S$ provides the greatest reach in 
$x_{gluon}$.  However, the reliability of fixed-order perturbative QCD as 
well as dominance of the $qg$ subprocess improve  
with greater $Q_T$.  The maximum value $Q_T \simeq $ 7.5 GeV attainable 
at $\sqrt S = 50$ GeV argues for a larger $\sqrt S$.    
  
Comparing the magnitudes of the prompt photon and massive lepton pair 
production cross sections~\cite{ref:BGKDYX}, we note that 
the inclusive prompt photon cross section is a factor of 1000 to 4000 greater 
than the massive lepton-pair cross section integrated over the mass interval 
2.0 $< Q <$ 3.0 GeV, depending on the value of $Q_T$.  This factor is 
attributable in large measure to the factor $\alpha_{em}/(3 \pi Q^2)$ 
associated with the decay of the virtual photon to $\mu^+ \mu^- $.  
Again taking $Ed^3\sigma/dp^3 = 10^{-3} \rm{pb/GeV}^2$ as the minimum 
accessible cross section, we establish that 
the real photon cross section may be measured to $p_T =$ 14, 33, and 52 GeV at 
$\sqrt S =$ 50, 200, and 500 GeV, respectively.  The corresponding reach in 
$x_T$ is $x_T = 2 p_T/\sqrt S =$ 0.56, 0.33, and 0.21 at $\sqrt S =$ 50, 200, 
and 500 GeV.  

The significantly smaller cross section in the case of massive lepton-pair 
production means that the reach in $x_{gluon}$ is restricted to a factor of 
about two to three less, depending on $\sqrt S$ and $Q$, than that potentially 
accessible with prompt photons in the same sample of data.  Nevertheless, it 
is valuable to be able to investigate the gluon density with a process that 
has reduced experimental and theoretical systematic uncertainties from those 
of the prompt photon case.  

In Ref.~\citen{ref:BGKDY} we compared our spin-averaged cross sections with 
available fixed-target and collider data on massive lepton-pair production at 
large values of $Q_T$, and we were able to establish that fixed-order 
perturbative calculations, without resummation, should be reliable for 
$Q_T > Q/2$.  The region of small $Q_T$ and the 
matching region of intermediate $Q_T$ are complicated by some level of 
phenomenological ambiguity.  Within the resummation approach, 
phenomenological non-perturbative functions play a key role in fixing the shape 
of the $Q_T$ spectrum at very small $Q_T$, and matching methods in the 
intermediate region are hardly unique.  For the goals we have in mind, it would 
appear best to restrict attention to the region $Q_T \geq Q/2$.

\section{Predictions for Spin Dependence}

Given theoretical expressions that relate the spin-dependent cross section 
at the hadron level to spin-dependent 
partonic hard-scattering matrix elements and polarized parton densities, we 
must adopt models for spin-dependent parton densities in order to obtain 
illustrative numerical expectations.  The current deep inelastic scattering 
data do not constrain the polarized gluon density tightly, and most groups 
present more than one plausible parametrization.  Gehrmann and Stirling 
(GS)~\cite{ref:GS} present three such parametrizations, labelled GSA, GSB, and 
GSC.   In the GSA and GSB sets, 
$\Delta G(x,\mu_o)$ is positive for all $x$, whereas in the GSC set 
$\Delta G(x,\mu_o)$ changes sign.  After evolution to $\mu_f^2 = 100$ GeV$^2$, 
$\Delta G(x,\mu_f)$ remains positive for essentially all $x$ in all three sets, 
but its magnitude is small in the GSB and GSC sets.  We use the three 
parametrizations suggested by Gehrmann and Stirling, and we verified that the 
positivity requirement
$\left | \Delta f^j_{h}(x,\mu_f)/f^j_{h}(x,\mu_f) \right | \le 1$ is
satisfied.  

Other sets of spin-dependent parton densities have been published, e.g., the 
set of Gl\"uck, Reya, Stratmann, and Vogelsang (GRSV).~\cite{ref:GRSV} \ The 
three GS 
parametrizations of the polarized gluon density span a range of possibilities 
that is very similar to that spanned by the four gluon densities of GRSV.  It 
is not our purpose to "test", or to suggest tests, of existing parametrizations 
all of which will have been modified substantially by the time data 
are available from RHIC.  Rather, we use the existing parametrizations as 
illustrative possibilities in order to 
estimate the range of magnitudes that might be expected for $A_{\rm LL}$ at 
RHIC energies and to gauge the sensitivity that measurements of spin 
dependence in massive lepton-pair production may offer.  The goal, after all, 
is to measure the polarized gluon density, not to test parametrizations.  

\begin{wrapfigure}{r}{6.6cm}
        \epsfxsize = 7 cm
        \centerline{\epsfbox{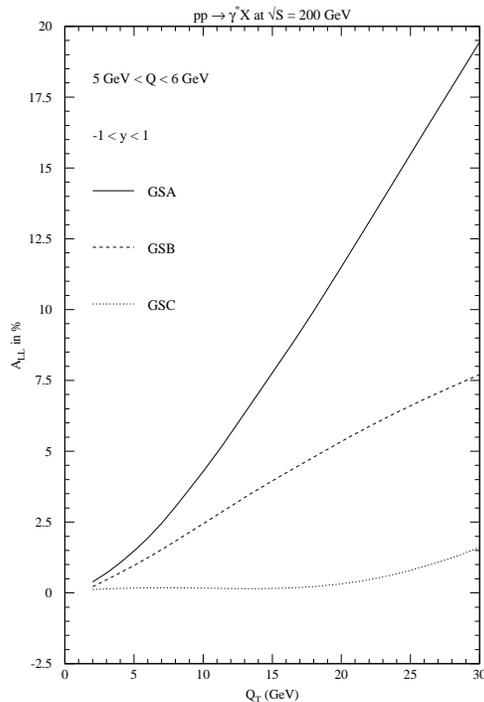}}
\caption{Computed longitudinal asymmetry $\protect A_{\rm LL}$ as a function 
of $Q_T$ for 
$p p \rightarrow \gamma^* X$ at $\sqrt S =$ 200 GeV.  The asymmetry is 
averaged over the rapidity interval -1.0 $< y <$ 1.0 and mass interval 
5.0 $<Q<$ 6.0 GeV, for three choices of the polarized parton densities, 
GSA, GSB, and GSC.}  
\label{fig:3}
\end{wrapfigure}
In this Section, we present two-spin longitudinal asymmetries for 
massive lepton-pair production as a function of transverse momentum.  Results 
are displayed for $pp$ collisions at the center-of-mass
energies $\sqrt{S}=$ 50, 200, and 500 GeV typical of the Brookhaven RHIC 
collider.

The two-spin longitudinal asymmetries, $A_{\rm LL}$, are 
computed in leading-order.  More specifically, we use leading-order 
spin-averaged and spin-dependent partonic subprocess cross sections, 
$\hat{\sigma}$ and $\Delta \hat{\sigma}$, with 
next-to-leading order spin-averaged and spin-dependent parton densities and a 
two-loop expression for $\alpha_s$.  The choice of a leading-order 
expression for $\Delta \hat{\sigma}$ is required because the 
full next-to-leading order derivation of $\Delta \hat{\sigma}$ has not been 
completed for massive lepton-pair production.  Experience with prompt photon 
production indicates that the leading-order and next-to-leading order results 
for the asymmetry are similar so long as both are dominated by the $qg$ 
subprocess.  To obtain the spin-dependent cross section, we use the GS 
polarized parton densities with the GS value $\Lambda^{(4)} = 231$ MeV.  This 
value of $\Lambda^{(4)}$ differs from the MRST value $\Lambda^{(4)} = 300$ 
MeV for the unpolarized densities used in our computation of the spin-averaged 
cross section.  The use of different values of $\Lambda^{(4)}$ for the 
spin-dependent and spin-averaged cross sections may appear unfortunate.  
However, there is not much of an alternative at present short of creating new 
sets of polarized parton densities based on the most up-to-date spin-averaged 
densities that we prefer to use for the spin-averaged cross section.  To change 
$\Lambda^{(4)}$ arbitrarily in the GS set to equal that of the MRST set would 
distort the spin-dependent densities.  

In Fig.~3, we present the two-spin longitudinal asymmetries,
$A_{\rm LL}$, as a function of  $Q_T$ for the three GS 
choices of the polarized gluon density.  The asymmetry becomes sizable for 
large enough $Q_T$ for the GSA and GSB parton sets but not in the GSC case.  
The asymmetry $A_{\rm LL}$ is nearly independent 
of the pair mass $Q$ as long as $Q_T$ is not too small.  This feature should 
be helpful for the accumulation of statistics; small bin-widths in mass 
are not necessary, but the $J/\psi$ and $\Upsilon$ resonance regions should 
be excluded.    

\begin{wrapfigure}{r}{6.6cm}
        \epsfxsize = 7 cm
        \centerline{\epsfbox{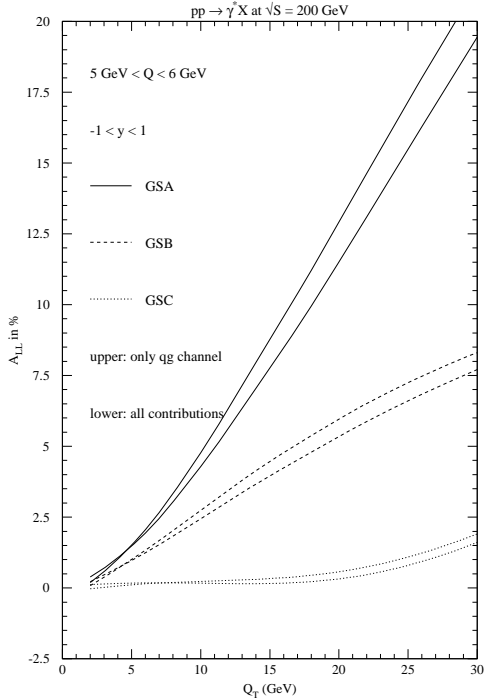}}
\caption{Comparison of the contribution of the $qg$ subprocess (upper curve) 
to the longitudinal asymmetry $\protect A_{\rm LL}$ with the total (lower curve) 
as a function of $Q_T$ for
$p p \rightarrow \gamma^* X$ at $\sqrt S =$ 200 GeV.  The asymmetry is 
averaged over the rapidity interval -1.0 $< y <$ 1.0 and over the interval 
5.0 $<Q<$ 6.0 GeV.  Results are shown for three sets of spin-dependent parton 
densities.}
\label{fig:4}
\end{wrapfigure}
As noted above the $qg$ subprocess dominates the {\it{spin-averaged}} cross 
section. It is interesting and important to inquire whether this dominance 
persists in the spin-dependent situation.  In Fig.~4, we compare 
the contribution to the asymmetry from the polarized $qg$ subprocess 
with the complete answer for all three sets of parton densities.  The $qg$ 
contribution is more positive than the full answer for values of $Q_T$ that 
are not too small; the 
full answer is reduced by the negative contribution from the 
$q \bar{q}$ subprocess for which the parton-level asymmetry 
${\hat{a}}_{\rm LL} = -1$.  At small $Q_T$, the net 
asymmetry may be driven negative by the $q \bar{q}$ contribution, and based on 
our experience with other calculations, from processes such as $gg$ that 
contribute in next-to-leading order.  For the GSA and GSB sets, we see that 
once it becomes sizable (e.g., 5\% or more), the total asymmetry from all 
subprocesses is dominated by the large contribution from the $qg$ subprocess. 

As a general rule in studies of polarization phenomena, many subprocesses can 
contribute small and conflicting asymmetries.  Asymmetries are readily 
interpretable only in situations where the basic dynamics is dominated by one 
major subprocess and the overall asymmetry is sufficiently large.  In the case 
of massive lepton-pair production that is the topic of this paper, when the 
overall asymmetry $A_{\rm LL}$ itself is small, the contribution from the 
$qg$ subprocess cannot be said to dominate the answer.  However, if 
a large asymmetry is measured, similar to that expected in the GSA case at the 
larger values of $Q_T$, Fig.~4 shows that the answer is dominated by 
the $qg$ contribution, and data will serve to 
constrain $\Delta G(x,\mu_f)$.  If $\Delta G(x,\mu_f)$ is small 
and a small asymmetry is measured, such as for the GSC parton set, or at small 
$Q_T$ for all parton sets, one will not be 
able to conclude which of the subprocesses is principally responsible, and no 
information could be adduced about $\Delta G(x,\mu_f)$, except that it 
is small.  

\begin{wrapfigure}{r}{6.6cm}
        \epsfxsize = 7 cm
        \centerline{\epsfbox{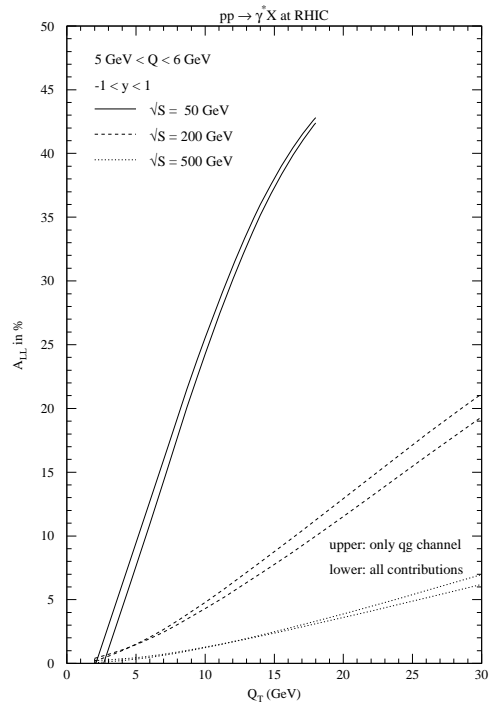}}
\caption{Computed longitudinal asymmetry $\protect A_{\rm LL}$ as a 
function of $Q_T$ for 
for $p p \rightarrow \gamma^* +X$ at $\protect\sqrt{S} =$
50, 200, and 500 GeV, averaged over the rapidity interval -1.0 $< y <$ 1.0 
and the mass interval 5.0 $<Q<$ 6.0 GeV.  Shown are both the complete 
answer at leading-order and the contribution from the $qg$ subprocess.  
The GSA set of polarized parton densities is used.}
\label{fig:5}
\end{wrapfigure}
In Fig.~5, we examine energy dependence.  For $Q_T$ not too small,  
$A_{\rm LL}$ in massive lepton pair production is well described by a 
scaling function of $x_T = 2Q_T/\sqrt S$, 
$A_{\rm LL}(\sqrt S,Q_T) \simeq h_{\gamma^*}(x_T)$.  
In our discussion of the spin-averaged cross sections, we took
$Ed^3\sigma/dp^3 = 10^{-3} \rm{pb/GeV}^2$ as the minimum accessible 
cross section.  Longitudinal 
asymmetries $A_{\rm LL} = 20 \%,\, 7.5 \%,\; \rm{and}\, 3 \%$ are predicted at 
this level of cross section at $\sqrt{S}=$ 50, 200, and 500 GeV when 
2 $< Q <$ 3 GeV, and $A_{\rm LL} = 11 \%,\, 5 \%,\, \rm{and}\; 2 \%$ 
when 5 $< Q <$ 6 GeV.  For a given 
value of $Q_T$, smaller values of $\sqrt S$ result in greater asymmetries 
because $\Delta G(x)/G(x)$ grows with $x$.  

The predicted cross sections 
in Fig.~2 and the predicted asymmetries in Fig.~5 should make it possible 
to optimize the choice of center-of-mass energy at which measurements might 
be carried out.  At $\sqrt{S}=$ 500 GeV, asymmetries are not appreciable in 
the interval of $Q_T$ in which event rates are appreciable.  At the other 
extreme, the choice of $\sqrt{S}=$ 50 GeV does not allow a sufficient range in 
$Q_T$.  Accelerator physics considerations favor higher energies since the 
instantaneous luminosity increases with $\sqrt{S}$.  Investigations in 
the energy interval $\sqrt{S}=$ 150 to 200 GeV would seem preferred.  
 
So long as $Q_T \ge Q$, the asymmetry in massive lepton-pair 
production is about the same size as that in prompt real photon production, 
as might be expected from the strong similarity of the production dynamics in 
the two cases.  As in massive lepton-pair production, $A_{\rm LL}$ in prompt 
photon production is well described by a scaling function of 
$x_T = 2p_T/\sqrt S$, $A_{\rm LL}(\sqrt S, p_T) \simeq h_{\gamma}(x_T)$.  
For $Ed^3\sigma/dp^3 = 10^{-3} \rm{pb/GeV}^2$, we predict 
longitudinal asymmetries $A_{\rm LL} = 31 \%,\, 17 \%,\, \rm{and}\; 10 \%$ in 
real prompt photon production at $\sqrt S =$ 50, 200, and 500 GeV.  

Although the $qg$ Compton subprocess is dominant, one might question whether
uncertainties associated with the quark density compromise the possibility to
determine the gluon density.  In this context, it is useful to
recall~\cite{ref:elbjwq} that when the Compton subprocess is dominant, the 
spin-averaged cross section may be rewritten in a form in 
which the quark densities do not appear explicitly:

\begin{equation}
 \frac{Ed^3\sigma_{h_1 h_2}^{l\bar{l}}}{dp^3} \approx
 \int dx_1 \int dx_2 \left( \frac{F_2(x_1,\mu_f^2)}{x_1} G(x_2,\mu_f^2)
 \frac{Ed^3\hat{\sigma}_{qg}^{l\bar{l}}}{dp^3} +
 (x_1 \leftrightarrow x_2) \right) .
 \label{dis1}
\end{equation}
Likewise, the spin-dependent cross section may be expressed 
as~\cite{ref:elbjwq}
\begin{equation}
 \frac{Ed^3\Delta \sigma_{h_1 h_2}^{l\bar{l}}}{dp^3} \approx
 \int dx_1 \int dx_2 \left( 2 g_1(x_1,\mu_f^2) \Delta G(x_2,\mu_f^2)
 \frac{Ed^3\Delta {\hat{\sigma}}_{qg}^{l\bar{l}}}{dp^3}
 + (x_1 \leftrightarrow x_2) \right).
 \label{dis2}
\end{equation}
In Eqs.~(\ref{dis1}) and~(\ref{dis2}), $F_2(x,\mu_f^2)$ and $g_1(x,\mu_f^2)$
are the proton structure functions {\it measured} in spin-averaged and
spin-dependent deep-inelastic lepton-proton scattering.  It is evident 
that the production of massive lepton-pairs at large enough $Q_T$ will
determine the gluon density provided the proton structure functions are
measured well in deep-inelastic lepton-proton scattering. 

\section{Discussion and Conclusions}

In this paper we summarize a calculation of the 
longitudinal spin-dependence of massive lepton-pair production at 
large values of transverse momentum.  We provide polarization 
asymmetries as functions of transverse momenta that may be 
useful for estimating the feasibility of measurements of spin-dependent cross 
sections in future experiments at RHIC collider energies. 
The Compton subprocess dominates the dynamics in longitudinally 
polarized proton-proton reactions as long as the polarized gluon density 
$\Delta G(x,\mu_f)$ is not too small.  As a result, two-spin measurements in 
massive lepton-pair production in polarized $pp$ scattering should 
constrain the size, {\it {sign}}, and Bjorken $x$ dependence of 
$\Delta G(x,\mu_f)$.  

Significant values of $A_{\rm LL}$ (i.e., greater than 5 \%) 
may be expected for $x_T = 2Q_T/ \sqrt S > 0.10$ if the polarized gluon density 
$\Delta G(x,\mu_f)$ is as large as that in the GSA set of polarized parton 
densities.  If so, the data could be used to determine the polarization of the 
gluon density in the nucleon.  On the other hand, for small $\Delta G(x,\mu_f)$, 
dominance of the $qg$ subprocess is lost, and $\Delta G(x,\mu_f)$ is 
inaccessible. 

Various methods have been discussed in the literature to access the 
spin-averaged and spin-dependent gluon densities.  These include 
inclusive/isolated prompt photon production at large transverse momentum 
(with or without a tagged recoil jet); the Drell-Yan process at large $Q_T$, 
as advocated in this paper; hadronic jet production at large $p_T$; and 
heavy flavor production such as charm ($c$) and bottom ($b$), thought to be 
mediated by the subprocess $g + g \rightarrow c + {\bar c} + X$.  Each has 
its possibilities and drawbacks.  Hadronic jet production benefits from the 
largest rate, but the large number of contributing subprocesses and the 
complications of jet definition mitigate against it.  Inclusive prompt 
photon production is theoretically clean but for the non-perturbative 
long-range fragmentation contribution.  However, experimenters measure 
{\it isolated} photons, and isolation renders the contact with theory 
somewhat murky.~\cite{ref:BGQ} \ The Drell-Yan process seems ideal, but its 
rate is relatively low.  Heavy flavor production is an enigma.  The charm quark 
is so light that calculations based on fixed-order perturbative QCD are of 
questionable reliability at collider energies.  The bottom quark seems to be 
heavy enough to justify perturbation theory, but why does the experimental cross 
section at Fermilab collider energies exceed next-to-leading order QCD 
predictions by a factor of 2 or 3?  

The dynamics of spin-dependence in hard-scattering processes is
a sufficiently complex topic, and its understanding at an early stage in its
development, that several defensible approaches for extracting polarized
parton densities deserve to be pursued with the expectation that consistent
results must emerge.  For the first time, RHIC offers an opportunity to 
explore the dynamics of spin dependence in hadron reactions at values of 
the hard scale that are sufficiently large that perturbative QCD ought to 
work.  It will take several years of experimentation and analysis for a 
clear and consistent picture to emerge of spin-dependent parton densities, 
just as it has in the study of spin-averaged parton densities.  The long 
term support of this program by the Brookhaven Laboratory and by the funding 
agencies will be essential for realization of the considerable potential 
that RHIC offers.  

\section*{Acknowledgments}
ELB is pleased to acknowledge the warm hospitality of the RIKEN Laboratory 
and of the Symposium hosts, particularly Shunzo Kumano, Masayasu Ishihara, 
Shunichi Kobayashi, Toshi-Aki Shibata, and Koichi Yazaki.  Work in the High 
Energy Physics Division at Argonne National Laboratory is supported by the 
U.S. Department of Energy, Division of High Energy Physics, 
Contract W-31-109-ENG-38.  This work was supported in part by DOE contract 
DE-AC05-84ER40150 under which Southeastern Universities Research Association 
operates the Thomas Jefferson National Accelerator Facility.  
MK is supported by Bundesministerium f\"ur Bildung
        und Forschung under Contract 05 HT9GUA 3, by Deutsche
        Forschungsgemeinschaft under Contract KL 1266/1-1, and by the
        European Commission under Contract ERBFMRXCT980194.

\end{document}